\pdfoutput=1
\documentclass[10pt,letterpaper]{article}
\usepackage[letterpaper,margin=1in,pdftex]{geometry}
\usepackage{abstract}
\usepackage{booktabs}
\usepackage{colortbl}
\usepackage{fancyhdr}
\usepackage{graphicx}
\usepackage{pifont}
\usepackage{subfig}
\usepackage{titlesec}
\usepackage{titling}
\usepackage{xcolor}
\usepackage[hidelinks]{hyperref}
\usepackage{breakurl}

\titleformat*{\section}{\Large\scshape\flushright}
\titleformat*{\subsection}{\large\scshape\flushright}
\newcommand{\KeyAuth}[0]{\textsc{KeyAuth}}
\newcommand{\subtitle}[0]{Bringing Public-key Authentication to the Masses}
\newcommand{\yes}{\ding{51}} % Makes a check mark

\begin{document}

\pretitle{\begin{flushright}\huge\scshape}
\title{KeyAuth\\\Large \subtitle{}}
\posttitle{\end{flushright}}

\preauthor{\begin{flushright}}
\author{\textsc{Travis Z. Suel}\\
\href{mailto:keyauthsoft@gmail.com}{keyauthsoft@gmail.com}}
\postauthor{\end{flushright}}

\date{}
\maketitle

\thispagestyle{empty}

\begin{abstract}
Passwords are a fragile, inadequate, and insecure tool for authenticating users,
and are especially fraught with problems when used to secure access to network
resources and services.  In many cases, passwords provide a false sense of
security.  Creating passwords which are both secure (i.e., hard for attackers to
guess) and easy for humans to remember is, at best, a paradoxical task because
these two criteria are diametrically opposed.  Fortunately, a far more secure
and user-friendly alternative is available.  Public-key cryptography provides a
means of both identifying and authenticating users without the need for
passwords.  \KeyAuth{} is a generic and universal implementation of public-key
authentication aimed at supplanting password-based authentication and
significantly improving the security of network accessible resources by
enhancing the usability of frequently used authentication mechanisms.
\KeyAuth{} is an application-, language-, operating system-, and
protocol-independent public-key authentication service.
\end{abstract}

\section*{The Problem}

\begin{center}
	\textbf{Authentication on the web is \emph{completely broken}.}
\end{center}

\begin{minipage}[t]{0.5\textwidth}
	\begin{itemize}
		\item Passwords are tedious to create.
		\item Passwords are tedious to use.
		\item Passwords are tedious to manage.
		\item Passwords unduly burden end-users.
	\end{itemize}
\end{minipage}
\begin{minipage}[t]{0.5\textwidth}
	\begin{itemize}
		\item Password databases are stolen regularly.
		\item Authentication is constantly reimplemented.
		\item All password amelioratives fall short or fail.
		\item Passwords ultimately degrade security.
	\end{itemize}
\end{minipage}

\begin{center}
	\textbf{One thing is strikingly clear: \emph{Passwords Are Not Working.}}
\end{center}

\section*{The Solution}

\begin{itemize}
	\item Fixing authentication on the web requires freeing users from
	      passwords.  How?
	\item Replace passwords with public-key authentication.
	\item SSL supports public-key authentication of users, but nobody uses this
	      facility. Why?
	\item Public-key authentication of users with SSL is not practical or
	      usable.
	\item We need a new system designed from the ground up for authenticating
	      users.
	\item This system must not burden the user with credential management.
\end{itemize}

\begin{center}
	\KeyAuth{} \textbf{is that new public-key authentication system built
	specifically for the end-user.}
\end{center}

\begin{figure}
	\centering
	\subfloat[Password management quickly becomes cumbersome for humans.]{\label{fig:passuse}\includegraphics[scale=1.0]{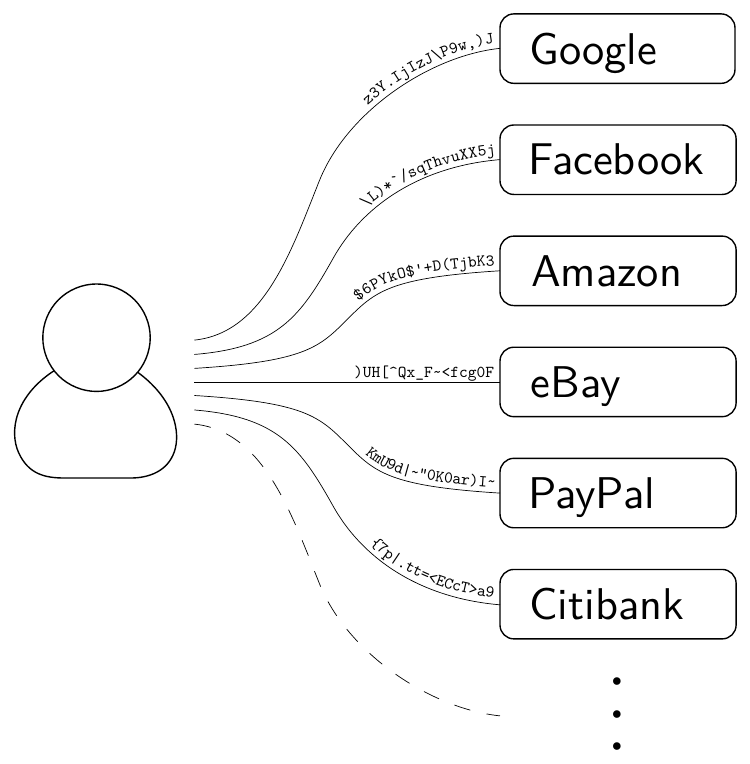}}~
	\subfloat[\KeyAuth{} allows a user to manage only one set of credentials.]{\label{fig:keyuse}\includegraphics[scale=1.0]{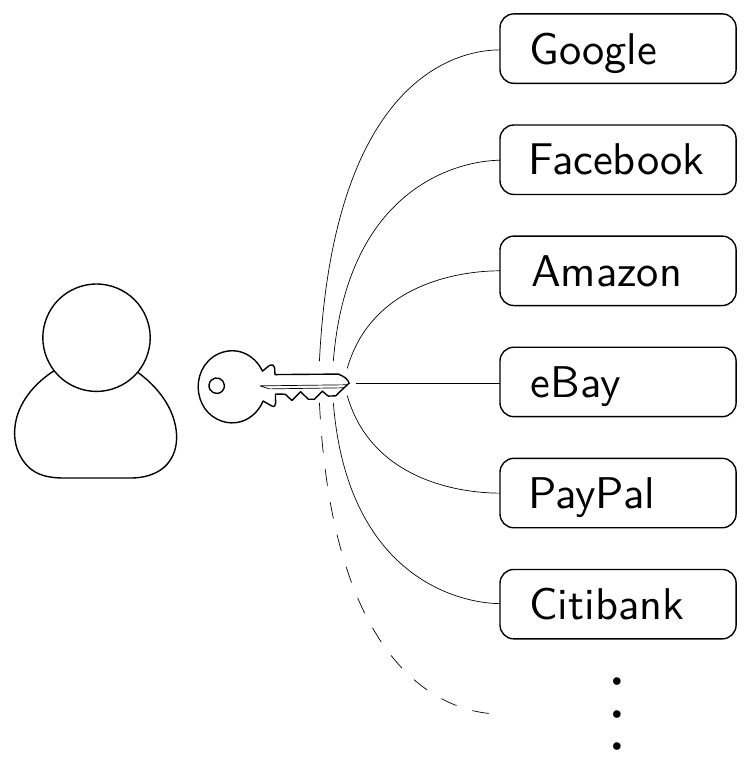}}
	\caption{Password authentication (left) vs. \KeyAuth{} (right).}
	\label{fig:pass_vs_pka}
\end{figure}

\section*{A Fundamentally Different Paradigm}

\KeyAuth{} is an authentication system which employs \emph{public-key
authentication} (PKA), an authentication paradigm fundamentally different from
password-based mechanisms, to verify the identity of users (see
Fig.~\ref{fig:pass_vs_pka}).  \KeyAuth{} solves the problem of credential reuse
allowing end-users to safely authenticate to any number of services with a
single set of credentials.  Requiring only one set of credentials significantly
eases the burden of management.  Users are no longer required to memorize long,
awkward, and random character strings solving the problems of password
management and usability.  In addition, \KeyAuth{} effectively eliminates the
error-prone generation of passwords.  Key pair generation is an automated
process which a user has to do only once.  Further, \KeyAuth{}, unlike current
PKA implementations, uniquely decouples the authentication process and logic
from individual applications and operating system facilities.  Instead
\KeyAuth{} provides public-key authentication as a lightweight, standalone
daemon which elides the hardware requirement (e.g., smart cards) of current PKA
systems eliminating significant barriers to the adoption of PKA by a broader
audience.  Providing a standalone service enables software developers to
properly factor authentication logic out of their applications allowing them to
devote more time and energy towards achieving design goals.  Therefore, as the
security of \KeyAuth{} improves, so does the security of client applications.
Finally, users and organizations do not have to rely on third-parties to perform
authentication on their behalf as they would with single sign-on services (e.g.,
those provided by Google and Facebook, see below).  Thus, \KeyAuth{} is a
complete replacement for password-based authentication that improves security
and usability and can be deployed in environments where smart cards are
impractical.  \KeyAuth{}

\begin{itemize}
	\item uses RSA public-key cryptography,
	\item uses OpenSSL to provide all cryptographic functionality,
	\item requires no specialized hardware,
	\item is application-, language-, operating system-, and protocol-independent,
	\item employs a simple and lightweight JSON protocol, and
	\item supports non-persistent connections (i.e., authentication over HTTPS).
\end{itemize}

\noindent Next we discuss the principles of public-key authentication and then
show that passwords are fundamentally flawed despite the many attempts to fix
them.  Table \ref{table:solutions} illustrates the many advantages of \KeyAuth{}
over passwords and the various password amelioratives such as password policies,
password safes, and locker services as well as single sign-on (all discussed
below).

\subsection*{Public-key Authentication}

\begin{table}
	\centering
	\caption{\KeyAuth{} represents a fundamentally different paradigm to
	         authentication with many benefits.} 
	\setlength{\aboverulesep}{0ex}
	\setlength{\belowrulesep}{0ex}
	\setlength{\extrarowheight}{0.65ex}
	\begin{tabular}{ r c c c >{\columncolor[gray]{0.85}}c }
	\toprule
	& Password policies & Password lockers & Single sign-on & \KeyAuth{} \\
	\midrule
	Improves usability               &  ---   & \yes{} &  \yes{} & \yes{} \\
	Simplifies management            &  ---   & \yes{} &  \yes{} & \yes{} \\
	Obviates generation              &  ---   &  ---   &  \yes{} & \yes{} \\
	Secures reuse                    &  ---   &  ---   & partial & \yes{} \\
	No third-party dependencies      & \yes{} &  ---   &   ---   & \yes{} \\
	Simplifies implementation        &  ---   &  ---   &   ---   & \yes{} \\
	Factors out authentication logic &  ---   &  ---   &   ---   & \yes{} \\
	\midrule
	Fundamentally different paradigm &  ---   &  ---   &   ---   & \yes{} \\
	\bottomrule
	\end{tabular}
	\label{table:solutions}
\end{table}

Public-key authentication utilizes \emph{public-key cryptography} (also referred
to as asymmetric cryptography) to identify and authenticate users, rendering
passwords obsolete.  In public-key cryptosystems, an end-user generates a
\emph{key pair}, using an easily automated process, composed of a \emph{public-}
and a \emph{private-key}.  Public-keys are used to transform information into an
unreadable, encrypted form, but they are incapable of reversing that
transformation.  Only private-keys can decrypt the results.  Furthermore,
private-keys cannot be derived from public-keys.  This property of public-key
cryptography allows for more robust and more secure authentication because PKA
does not depend on users memorizing passwords.  The criteria of usability and
security do not compete in PKA systems as in password-based schemes.

In addition, the nature of public-key cryptosystems does not require
authentication systems to store sensitive information (e.g., passwords) which
could be used to gain unauthorized access.  The only information necessary to
verify the identity of an end-user is the public-key of that user.  Should an
attacker successfully obtain a public-key through an unpatched security
vulnerability, they would be unable to use that public-key to gain further
unauthorized access.  Since a public-key can only be used to verify an
authorized user's identity, it cannot be used to pose as an authorized user.
PKA systems require neither the transmission of a user's private-key, nor its
storage in a central location meaning that reusing a key pair, unlike reusing a
password, is not fundamentally insecure.  Quite the contrary, key reuse is a
fully supported use case of properly designed public-key cryptosystems.
Potential attackers are also prevented from obtaining valid recoverable
credentials from a single intrusion.  These aspects of public-key authentication
make \KeyAuth{} far more secure and fault-tolerant than passwords, even in the
presence of inevitable bugs, security vulnerabilities, and successful breaches.

Public-key authentication is a time-tested and robust approach to user
authentication.  However, all popular implementations are directly tied to
specific applications, operating systems, protocols, or specialized hardware.
Transport Layer Security (TLS) and Secure Sockets Layer (SSL) provide PKA
facilities.  However, this functionality is an extension of identifying web
servers and imposes restrictions which, while useful for identifying network
nodes, render it impractical for identifying people.  When using TLS/SSL to
authenticate users, each user's public-key must have an associated certificate
provided by a certificate authority.  Also, users have to install their key
pairs in their browsers which associates public-keys with browser installations,
not users.  Moreover, this approach limits such authentication to web
applications served over HTTPS.  Traditional OS-level PKA requires every valid
user to have an operating system account on the computer performing
authentication and is intended for use with specialized hardware (e.g., smart
card readers).  Card readers are crucial components for local authentication in
secured environments such as government offices, but they serve as an impediment
to mainstream adoption as typical end-users do not have access to such hardware
for personal use.  Unlike current PKA implementations, \KeyAuth{} disentangles
the authentication logic from specific applications, operating systems, and
protocols making public-key authentication a practical replacement for
passwords.

\section*{Passwords are the Problem}

For a password to be secure it must be difficult for both a human and a computer
to guess, yet easy to remember.  This security is typically achieved by creating
passwords as strings of random characters which include letters of both cases,
numbers, and various other symbols, including punctuation and brackets
\cite{arsTechnica:gallagher2011}.  As the length of these random amalgams of
characters increase, so does the difficulty of guessing them.  However, as
password complexity and length increases, passwords become more challenging for
humans to memorize and type; there is a steep drop in usability because
passwords are not compatible with the mechanisms of human memory.  Human memory
employs a strategy called \emph{chunking} in which an individual improves recall
by dividing information into chunks, such as words or syllables,  which are
perceived to have meaning \cite{wikipedia:chunking}.  Unfortunately, strings of
random characters do not easily decompose into retainable chunks.  Thus,
choosing between password security and usability is a zero-sum exercise.
Passwords are an inherently ineffective credential mechanism because they rely
directly on human memory \cite{arsTechnica:goodin2012_passphrase,
wired:mcmillan2012}.

Implementing and deploying password-based authentication is a common source of
security vulnerabilities.  Any such system must store user passwords in some
form.  Because storing passwords in plaintext is a security vulnerability,
passwords are often stored using a one-way hash algorithm (e.g., MD5 or SHA1).
However, this is often done improperly without a \emph{salt}\footnote{A string
of random characters appended to a password before hashing to defend against
dictionary attacks.} or with a salt common to all users.  Regardless of how
securely passwords are stored, maintaining them in a central location
conveniently provides an attacker with a single point of access.  If an attacker
discovers a means of unauthorized access to this highly sensitive information
(possibly by exploiting a SQL injection vulnerability), they can mount an
offline dictionary attack and likely recover passwords from a subset of users
and gain further unauthorized access.  Even software which properly salts
passwords before hashing is vulnerable because of password reuse.  An attacker
who gains unauthorized access to an insecure but non-critical system, perhaps a
public-facing content management system, can reasonably expect some recovered
passwords to be valid with other more secure and high-value systems, such as
source code repositories or e-mail servers \cite{arsTechnica:bright2011}.  This
attack vector can be effective both within and across organizational boundaries.

Attack vectors arise from password reuse because most organizations and software
implement their own authentication logic.  Continually re-implementing
authentication logic (within and between organizations) not only wastes
productivity but also increases the number of systems with potentially
exploitable security flaws which expands attack surfaces.  An attacker can
exploit such flaws in a vulnerable system to recover passwords and subsequently
access otherwise inaccessible secure systems.  There are no compelling reasons
why the software industry should take such a haphazard approach to implementing
authentication logic.  Rather, this effort should be devoted to achieving the
unique design goals of individual applications.  Moreover, the plurality of
disjoint implementations means as bugs are fixed in one application, similar or
identical bugs may go unnoticed and unpatched in others.  As a result, passwords
and the common approaches taken to implementing them are irredeemably flawed
approaches to the problem of authenticating users.  In summary, password
vulnerabilities are the result of many factors:

\[
\textit{password vulnerabilities} = \left\{
	\textit{\parbox{11ex}{\centering password reuse}} \times
	\textit{\parbox{13ex}{\centering centralized datastore}} \times
	\textit{\parbox{08ex}{\centering logic errors}} \times
	\textit{\parbox{19ex}{\centering abundance of implementations}}
\right\}
\]

\section*{Passwords Cannot Be Fixed}

\subsection*{Password Policies}

Many organizations attempt to improve the security of passwords and prevent
reuse by imposing \emph{password policies} on users which attempt to promote
secure behavior.  However, password policies are often examples of
\emph{anti-patterns}\footnote{A solution which is commonly used in software
development, but which is either ineffective or counterproductive.} and suffer
from serious unforeseen consequences \cite{arsTechnica:gallagher2011}.  These
policies seek to improve password security by requiring users to create highly
complex passwords of a secure length, and, in many cases, requiring users to
change passwords on a regular basis.  These policies, while well-intentioned,
are frequently overzealous in their complexity and, thus, unrealistic for users
to abide.  Ultimately, password policies compel users to store passwords in
plaintext, often by writing them down, and compromise their security.  Users
often have no other choice as they cannot memorize a constant stream of new,
complex passwords.  Password policies also increase IT support costs as more
users are required to call for help when they invariably forget their new,
secure password \cite{arsTechnica:gallagher2011}.

A single password policy is composed of multiple requirements such as minimum
length, required character types, and password expiration.  Compounding this
problem further, many organizations are subject, internally, to multiple
password policies.  This makes creating new passwords difficult because these
various policies are often incompatible and failing to post clear and explicit
password requirements is confoundingly common.  One password policy may prohibit
a specific set of characters while another policy may ban all non-alphanumeric
characters.  Still another policy may impose an arbitrary maximum length.  As an
example of the complexities such policies introduce, performing a Google search
for ``\texttt{amazon password policy}'' yields no helpful results from Amazon.
When changing an Amazon password, if a user enters an invalid password (perhaps
containing disallowed characters), Amazon silently rejects the password leading
the user to believe it has been accepted.  Password policies struggle to treat
the symptoms of passwords while further complicating their use and ignoring the
root of the problem.

\subsection*{Password Safes}

To cope with the ever-increasing number and complexity of passwords, some users
choose to employ \emph{password safes} or \emph{lockers} to manage their myriad
credentials.  A password safe can either manage a locally stored, usually
encrypted, database, or it can manage a database stored in a remotely accessible
location.  Password safes which manage a local database (e.g., KeePass) help to
alleviate the user of remembering long complex passwords without resorting to
plaintext storage.  However, if a user owns multiple devices (e.g., a laptop, a
tablet, and a smartphone), they must synchronize the database on each device
every time they add, update, or delete a password entry.  Password locker
services (e.g., LastPass) relieve the user of password database management by
storing the database ``in the cloud'' making it easily accessible to multiple
devices.

Password safes, while mitigating poor usability and cumbersome management, do
not fix passwords.  Most passwords lockers are still secured using a password,
and locker services exacerbate the aspect of centralized storage since passwords
to multiple services are collected from many users.  Safes and lockers also do
nothing to solve the problem of generation and only partially remedy reuse.
Creating a new account or changing the password on an existing account still
requires a person to know the password requirements of a particular service.
Further, locker services also make people dependent on third-parties to store
and maintain their passwords.  It is neither unheard of nor unreasonable for
industry heavyweights, including Google and Microsoft, to suffer the occasional
service outage.  Outages of Gmail or Office 365 may be a nuisance or significant
interruption, but, if a locker service suffers an outage, their users are
effectively locked out of all other services since access to all credentials has
been temporarily suspended.  Lastly, though importantly, password safes only
prevent the fallout of reuse for people who chose to use them.  Businesses and
organizations are still subject to the security implications of password reuse
by those who do not.

\subsection*{Single Sign-on}

\emph{Single sign-on} (SSO) systems only partially mitigate the fundamental
flaws of passwords.  These systems, ultimately, still attempt to derive their
security from passwords, and users are still required to authenticate with SSO
services using passwords.  If a password or SSO service is comprised, all
systems which rely on that SSO service are suddenly susceptible to security
breaches.  Moreover, using the SSO service of another organization introduces
third-party dependencies.  Supplementary to local accounts, all users need
accounts with the SSO provider, and an organization must forfeit some level of
control to that provider.  A provider may choose to alter their protocol in a
way that is not backwards compatible or even discontinue service entirely. The
client organization may have little recourse but to bear the cost of retooling
their software or migrating to another, possibly incompatible, service.  A
client organization may inadvertently run afoul of the business policies of
their SSO provider and have their access revoked without prior notice, a
scenario in which Grooveshark recently found themselves
\cite{arsTechnica:geuss2012}.  Finally, SSO systems introduce a great deal of
complexity to the authentication process rendering them difficult to implement
properly and securely~\cite{arsTechnica:goodin2012_sso}.  Some SSO flaws have
even been found to allow attackers unauthorized access without obtaining any
passwords.  Compounding the intricacies of SSO, many organizations are required
to support multiple SSO providers because their users do not share the same
provider.  While single sign-on clearly minimizes certain undesirable aspects of
passwords, it decidedly exacerbates others.

\section*{Conclusion and Further Reading}

Usability is the key to promoting secure behavior in end-users.  Make security
an easy and convenient habit and users will adopt secure conduct.  Passwords are
insecure because they burden users with extraneous and repetitive work which
unintentionally fosters bad behavior.  Most password amelioratives do solve some
of the shortcomings of passwords but ignore or even aggravate others.  Some of
these purported solutions even introduce new problems.  Unfortunately, none of
the amelioratives are capable of solving all the problems inherent to
password-based authentication (see Table \ref{table:solutions}).  \KeyAuth{}
bypasses these problems entirely by obviating the need for passwords with a
fundamentally different approach to user authentication and, therefore, relieves
people of all the associated tedious and error-prone management and overhead of
passwords.  The onus of herding the mundane and tiresome details of credential
security shifts from humans to software.  With \KeyAuth{}, instead of being
arduous, frustrating, and time consuming, proper security becomes innate,
intrinsic, and natural.

\renewcommand\refname{\vspace{-2em}}
\bibliographystyle{acm}
\bibliography{keyauth}

\end{document}